\documentclass[aps,superscriptaddress,reprint]{revtex4-1}
\usepackage{bm}
\usepackage{amsmath,amssymb,cases} 
\usepackage{graphicx}
\usepackage{xcolor}
\usepackage[normalem]{ulem}
\usepackage{braket}

\begin{document}

\title{Spatial Bloch oscillations of a quantum gas in a ``beat-note"  superlattice}
\author{L. Masi}
\affiliation{CNR Istituto Nazionale Ottica, 50019 Sesto Fiorentino, Italy}
\author{T. Petrucciani}
\affiliation{\mbox{European Laboratory for Nonlinear Spectroscopy (LENS), 50019 Sesto Fiorentino, Italy}}
\author{G. Ferioli}
\affiliation{CNR Istituto Nazionale Ottica, 50019 Sesto Fiorentino, Italy}
\author{G. Semeghini}
\affiliation{CNR Istituto Nazionale Ottica, 50019 Sesto Fiorentino, Italy}
\author{G. Modugno}
\affiliation{CNR Istituto Nazionale Ottica, 50019 Sesto Fiorentino, Italy}
\affiliation{\mbox{European Laboratory for Nonlinear Spectroscopy (LENS), 50019 Sesto Fiorentino, Italy}}
\affiliation{Department of Physics and Astronomy, University of Florence, 50019 Sesto Fiorentino, Italy.}
\author{M. Inguscio}
\affiliation{CNR Istituto Nazionale Ottica, 50019 Sesto Fiorentino, Italy}
\affiliation{\mbox{European Laboratory for Nonlinear Spectroscopy (LENS), 50019 Sesto Fiorentino, Italy}}
\affiliation{Department of Physics and Astronomy, University of Florence, 50019 Sesto Fiorentino, Italy.}
\author{M. Fattori}
\affiliation{CNR Istituto Nazionale Ottica, 50019 Sesto Fiorentino, Italy}
\affiliation{\mbox{European Laboratory for Nonlinear Spectroscopy (LENS), 50019 Sesto Fiorentino, Italy}}
\affiliation{Department of Physics and Astronomy, University of Florence, 50019 Sesto Fiorentino, Italy.}

\date{\today}

\begin{abstract}
We report the experimental realization of a new kind of optical lattice for ultra-cold atoms where arbitrarily large separation between the sites can be achieved without renouncing to the stability of ordinary lattices. Two collinear lasers, with slightly different commensurate wavelengths and retroreflected on a mirror, generate a superlattice potential with a periodic ``beat-note" profile where the regions with large amplitude modulation provide the effective potential minima for the atoms. To prove the analogy with a standard large spacing optical lattice we study Bloch oscillations of a Bose Einstein condensate with negligible interactions in the presence of a small force. The observed dynamics between sites separated by ten microns for times exceeding one second proves the high stability of the potential. This novel lattice is the ideal candidate for the coherent manipulation of atomic samples at large spatial separations and might find direct application in atom-based technologies like trapped atom interferometers and quantum simulators.    

\end{abstract}

\maketitle
Optical lattices are powerful tools to study and manipulate ultra-cold gases \cite{morsch2006dynamics}. They are used to investigate superfluidity in quantum gases \cite{PhysRevA.64.061603, PhysRevLett.86.4447}, to study non-linear dynamics of matter waves \cite{PhysRevLett.91.060402}, to perform quantum simulation of solid-state physics models \cite{gross2017quantum} and to operate atom interferometers for high precision measurements of forces \cite{Peters_Nature} and fundamental constants \cite{rosi2014precision, clade2006determination}. The optical interference on which optical lattices are based determines their key stability properties. Optical lattices created by retro-reflecting a laser beam of wavelength $\lambda$ on a single mirror are particularly appealing since the lattice period is exactly $\lambda /2$, is strongly immune to beam pointing instabilities \cite{Hosten1552} and the phase noise can be suppressed by stabilizing the mirror motion \cite{Peters_Nature}. In addition, the residual intensity noise of optical lattices is normally not an issue, since it induces common-mode fluctuations of the site potentials.  
The spatial periodicity of optical lattices based on counterpropagating beams is limited to the range from 0.2 to 0.7 $\mu$m with only few exceptions \cite{PhysRevA.62.051801}, mainly due to the available narrow-linewidth laser sources. There is however a strong interest in creating periodic potentials with larger with separations between the different sites for a variety of applications ranging from trapped-atom interferometry for the measurement of elecromagnetic fields, gravity and inertial forces \cite{PhysRevLett.118.230403}, atomtronics \cite{Amico_2017}, quantum simulation of Hubbard-like models \cite{PhysRevA.92.021402} and quantum computing \cite{Henriet2020quantumcomputing,PhysRevLett.123.170503}. This has led to the realization of different kinds of trapping potentials using acousto-optic deflectors \cite{Shin}, spatial light modulators \cite{Zupancic:16}, laser beams crossing at small angles \cite{PhysRevLett.95.010402} and magnetic traps \cite{Tiecke_2003}. However none of these methods can offer the stability of optical lattices realized with counterpropagating laser beams.
\\\\
In this work we demonstrate an innovative, large-spacing optical superlattice based on the beating note between two retroreflected optical  lattices with slightly different wavelengths $\lambda_1$ and $\lambda_2$, i.e. $|\lambda_1 - \lambda_2| < \lambda_{1,2}$. We show that, when the two wavelengths fullfil the condition $n\lambda_1 = (n+1)\lambda_2$ with $n$ integer, the resulting potential is periodic and, for sufficiently low lattice depths, the energy spectrum  of the superlattice is equal to the one of an optical lattice with wavelength $n \lambda_1$. Working with $n=20$ and standard laser wavelengths around $1 \mu$m, we realize an effective lattice period around $10 \mu$m. Our ``beat-note" superlattice (BNSL) is analog to the well-known superlattices that have revolutionized the field of semiconductors \cite{Esaki}, allowing to create effective lattice periods one order of magnitude larger than the intrinsic one \cite{Leo}. Similar approaches have been used also in optics using multilayer dielectric structures to observe traport phenomena typical of charged particles using light waves \cite{PhysRevLett.91.263902, PhysRevLett.94.127401}. In quantum gases experiments second harmonic superlattices, i.e. for $n=1$, have been used to create arrays of double well potentials \cite{Folling}.   
 
To investigate the properties of the BNSL, we prepare a Bose-Einstein condensate in its ground state, we employ it to measure the energy gaps between the first three bands and we study Bloch oscillations in the presence of an external force. These measurements demonstrate that the BNSL is equivalent to a standard lattice up to depths of the order of the recoil energy of the two combined lattices. Moreover, when cancelling the interatomic interactions by means of a magnetic Feshbach resonance \cite{d2007feshbach}, the dynamics exhibits a coherence up to 1 s, demonstrating how this new techinique provides very stable potentials with an arbitrarily long periodicity. 
\begin{figure}[t]
{\includegraphics[width=\columnwidth]{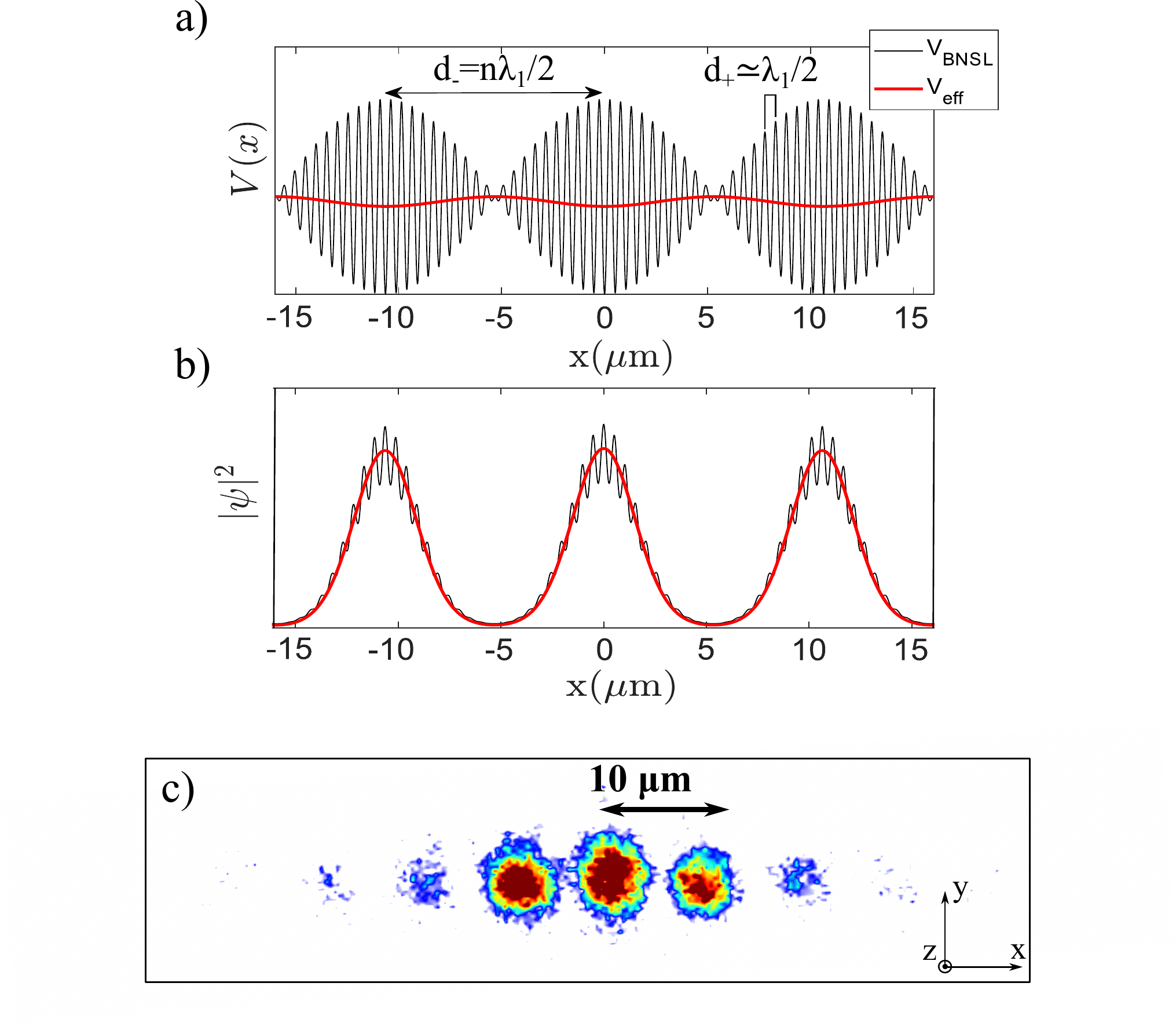}}
\caption{a) Plot of the beat-note optical lattice (thin line) and the correspondent effective potential  $V_{eff}$ (thick line). b) Profile of the ground-state atomic wavefunction in presence of a BNSL with a depth $V_0 =0.5 E_{R^{+}}$ (thin line) and in presence of a standard large spacing optical lattice with a depth equal to the effective depth of the BNSL (thick line). c) Density distribution of a non interacting condensate in the ground state of the BNSL that shows the spatial modulation with a period of 10 $\mu$m.}
\label{fig1}
\end{figure}\\
\begin{figure}[t]
{\includegraphics[width=\columnwidth]{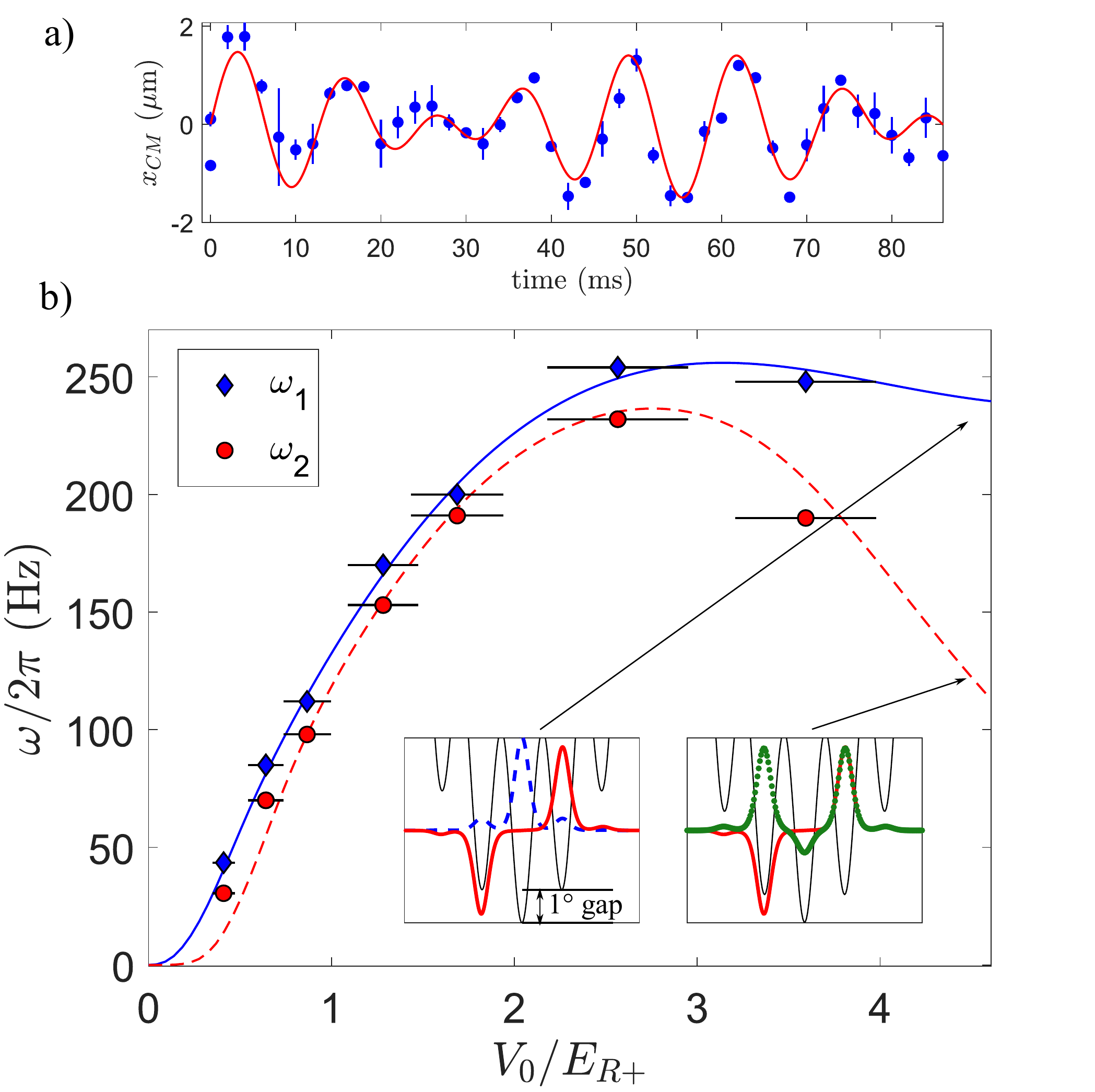}}
\caption{a) Center of mass oscillation of a BEC trapped in a single site of the BNSL. The evolution is the result of the beating between the frequencies related to the first two energy gaps of the spectrum $\omega_1$ and $\omega_2$. The line is a fit to the data. b) Measured values of $\omega_1$ and $\omega_2$ as a function of $V_0$ compared with the theoretical values, line and dasehd line respectively. Error bars on the lattice depth take into account the uncertainties of the beam size on the atoms and of the optical power. The insets show the BNSL potential around the minimum and compare two by two the wavefunctions of the first three on-site energy eigenstates for $V_0=4E_{R+}$, using dashed, solid and dotted lines respectively.}
\label{fig2}
\end{figure}\\
The superposition of two standing waves with the same optical depth $V_0$ and with a relative phase $\phi$ provides a trapping potential for the atoms
\begin{equation}
V(x)=V_0[1-\cos(k_-x  - \phi )\cos(2k_+x + \phi )]
\label{BL}
\end{equation}
with $k_+=(k_1+k_2)/2$, $k_-=k_2-k_1$ and $k_{1,2}=2\pi/\lambda_{1,2}$ \cite{SI}. The potential is the sum of a constant term $V_0$ plus a sinusoidal one with a fast spatial modulation of periodicity $d_+=\pi/k_+$ and with an amplitude varying over a distance $d_-=\pi/k_-$. When the two wavelengths fullfil the commensurability condition $n\lambda_1=(n+1)\lambda_2$, the potential is periodic over a distance $d_-=n \lambda_1/2$ realizing a BNSL (see Fig. (\ref{fig1}a)).
As pointed out in \cite{PhysRevA.99.043608}, a quantum particle evolving in a potential with a periodic spatial modulation experiences an attractive effect in comparison to a constant one with the same average value. 
The reason is that, although the modulation naturally increases the kinetic energy due to a coupling to high momentum states, the resultant modulation of the wavefunction, with maxima localized at the minima of the trap, causes a stronger reduction of the potential energy. In the perturbative limit, i.e. $V_0<<E_{R^+}$ with $E_{R^+}=\hbar^2 k_+^2/2m$, this attractive effect can be quantified with a negative potential equal to the square of the modulation amplitude divided by 8$E_{R^+}$ \cite{ashcroft1976solid,SI}. As a consequence we can approximate Eq.(\ref{BL}) with an effective potential  
\begin{equation}
V_{eff}(x)=V_0 -\frac{V_0^2 \cos^2(k_-x - \phi)}{8E_{R^+}}
\label{delta}
\end{equation}that is equivalent to the one of an optical lattice with periodicity $d_-$. The low $V_0$ limit implies that the tunneling between local minima of the potential in Eq.(\ref{BL}) is larger than the on-site energies such that the wavefunction is minimally affected by the fast spatial modulation of the trap (see Fig. 1b). This is no longer true for $V_0\simeq E_R+$ where the perturbative approach of Eq. (\ref{delta}) cannot be applied. However the BNSL continues to behaves like a single wavelength lattice if we restrict the analogy to the first band. Note that, in this regime of lattice depths, if the separation between the BNSL sites is sufficiently large, the effective tunneling between sites is so small to allow the loading of an array of independent atomic clouds (see \cite{SI}). Finally, for $V_0>>E_{R+}$ the potential deveopls its own peculiar spectrum and the analogy with a large spacing optical lattice is completely lost.


\color{black} In order to study the behavior of an ultra-cold gas in a BNSL we exploit an experimental setup similar to the one described elsewhere \cite{landini2012direct}. We work with a Bose-Einstein condensate of $^{39}K$ in the state $\ket{1,1}$ where the interatomic interactions can be tuned close to zero exploiting a broad magnetic Feshbach resonance  at 400 G. Once we have evaporatively cooled the gas to condensation, we adiabatically ramp up two optical lattices along the \textit{x} axis (see Fig. (\ref{fig1}c)) with wavelengths $\lambda_1=1064.5$ nm and $\lambda_2=1013.7$ nm.  A dipole trap beam that propagates along the same axis provides a radial harmonic potential with $\omega_{y,z} \approx 2\pi\cdot 200$  Hz. \color{black} The two lattices provide a BNSL with $d_+ \simeq 0.5$ $\mu$m and $d_-=10.6$ $\mu$m with $n=20$. Since the position of the effective minima depends on the relative phase $\phi$ between the two combined lattices, both the laser frequencies are locked to the same reference cavity with a relative stability of the order of ten kHz and tuned to provide $\phi=0$ \cite{SI}. We finally detect the in-situ atomic density from the orthogonal z direction (see Fig. (\ref{fig1}c)), ramping up the optical lattices to a a final depth $V_0 \sim 200 $ nK $\approx  E_{R^+} \sim 220 $ nK. The observed spatial separation of 10 $\mu$m between independent condensates localized in the effective sites of the BNSL confirms our expectations. \color{black} 

In our system we cannot calibrate the lattice depth using standard techniques based on atomic diffraction, since the coherence length of the condensate is smaller than $d_-$.
We therefore use a different method based on the study of the oscillation frequency of the condensate in a single site of the BNSL, in regimes where the tunneling energy $J_{eff}$ to neighbouring sites is negligible. \color{black} The dynamics is simply triggered by a sudden shift of the minima of the potential. A typical measurement result is shown in Fig.\ref{fig2}a) where we intentionally shift the minima by an amount sufficiently large to project the initial condensate wavefunction on the first three bands. The center of mass oscillation of the gas is then characterized by the beating of two frequencies that correspond to the first two energy gaps. In Fig.\ref{fig2}b) we compare the measured values with the theoretical expectations. If $V_0 \lesssim E_{R^+}$, according to Eq. (\ref{delta}), the effective potential is sinusoidal and its anharmonicity leads to two slightly distinct frequencies that scale with the square root of $V_{eff}$, i.e. linearly with $V_0$. For larger values of $V_0$, where the analogy with a large spacing standard lattice is no longer valid, the first energy gap of the BNSL deviates from the linear behaviour, while the second one reaches a maximum and then decreases to zero. In the left inset of Fig.\ref{fig2}b) we understand that the first gap becomes asymptotically equal to the potential shift of the two sites adiacent to the one with the lowest energy. The reduction of the second gap instead is due to the negligible energy difference between the antisymmetric and symmetric states of the second and third band respectively, as shown in the right inset of Fig.\ref{fig2}b).    
\begin{figure}[t]
\centerline{\includegraphics[width=1\columnwidth]{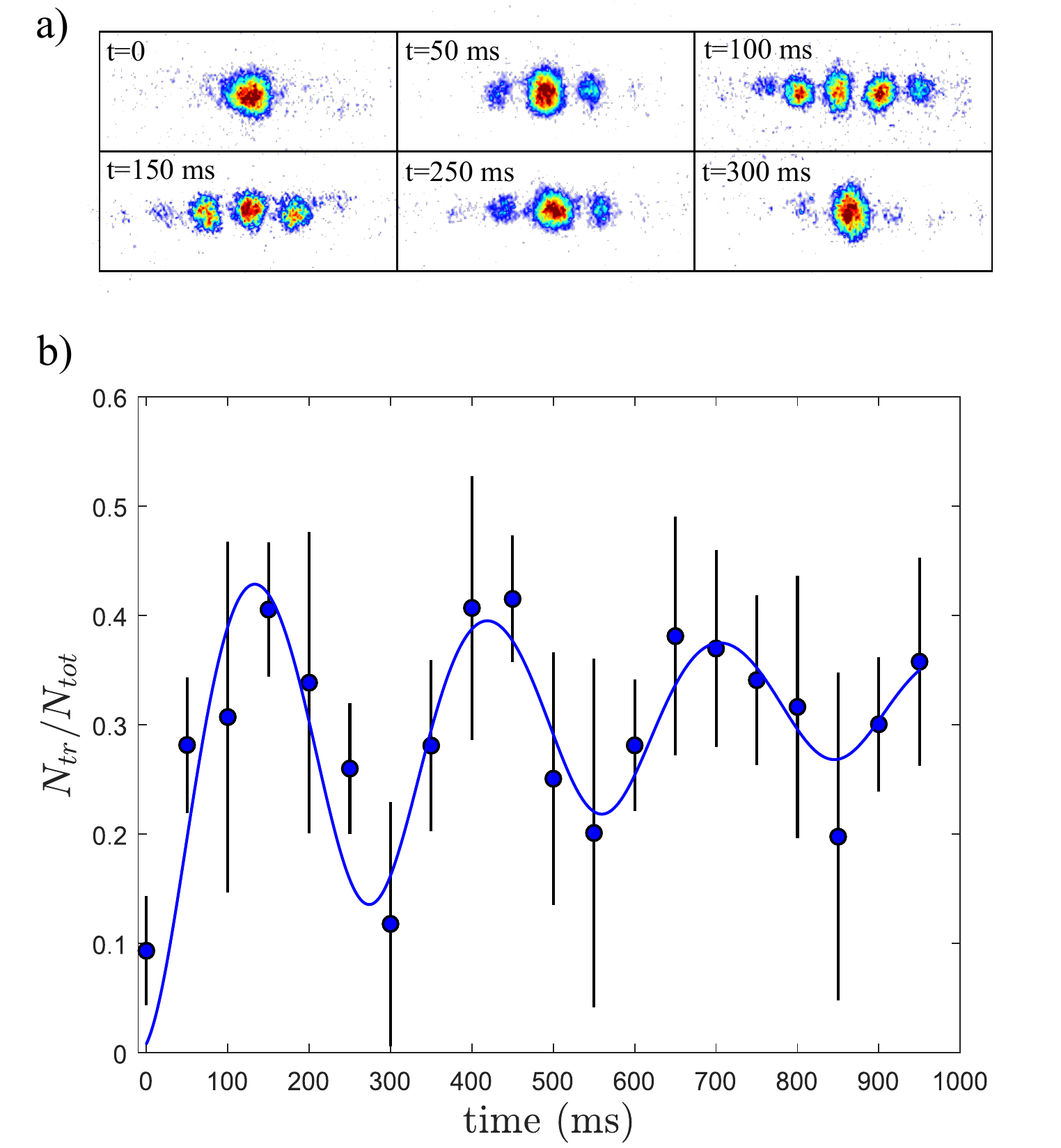}}
\caption{a) Absorption images of the BEC during a spatial Bloch oscillation in presence of an external force, characterized by a breathing motion of the width. In b) we report the time evolution of fractional atom number in the starting site $N_0$ and the sum of the populations in the neighbouring sites $N_{tr}$. The solid line is a fit perfomed using Eq. (\ref{bessel}) times an additional exponential decay of the amplitude. The error bars represent the statistical uncertainty and correspond to the standard deviation.}
\label{fig3}
\end{figure}\\
In order to prove the analogy between a BNSL and a single wavelength large spacing optical lattice in the low $V_0$ limit, we finally need to measure $J_{eff}$ and observe the coherent dynamics between its different sites \cite{Preiss, PhysRevLett.120.213201}. We then perform Bloch oscillations of a condensate in the BNSL in presence of an external force and detect the in trap atomic density evolution. Experimentally, we initially load a condensate of $N\simeq5\cdot10^3$ atoms in a single site of the BNSL with an initial depth $V_0 \approx$ 300 nK, that ensures a negligible tunneling between neighbouring sites. A spurious external magnetic field gradient that causes an acceleration $\approx 10^{-2}$g, where g=$9.81$m/s$^2$, is compensated by the optical force provided by the dipole trap propagating along the \textit{x} axis whose focus position is shifted with respect to the condensate by a distance equal to the Rayleigh range of the beam. In this way, by tuning the intensity of the dipole trap, we can finely adjust the total external force $F$ around zero, both in magnitude and sign, over a range of $\pm 5 \cdot 10^{-4}$ \textit{mg} \cite{SI}.


Once set the working value of the force, we switch off the dipole trap that provides confinement along x and we decrease the BNSL depth to $V_0=70(\pm2)$ nK. The procedure is performed adiabatically with respect to the trapping frequencies of a single site but on a time-scale much shorter then the Bloch period. During the oscillation we set the scattering length to a negligible value in order to minimize the decoherence induced by interactions. In Fig.\ref{fig3}a) we report examples of absorption images of the condensate during one Bloch oscillation. Since at $t=0$ all the atoms are localized in a single site, the center of mass motion is frozen and we observe only a symmetric breathing evolution of the size \cite{preiss2015strongly}. The on-site fractional populations evolve in time as \cite{hartmann2004dynamics}:
\begin{equation}
n_l(t)=\Bigr|\mathcal{J}_l\Bigr(\frac{4J_{eff}}{\delta}\sin{\Bigr(\frac{\delta t}{\hbar}\Bigr)}\Bigr)\Bigr|^2
\label{bessel}
\end{equation}
where $\mathcal{J}_l$ are Bessel functions of the first kind and $\delta=Fd_-$ is the energy difference between neighbouring sites. Experimentally we count the atoms that remain in the starting well $N_0$ and the ones that tunnel to the other sites $N_{tr}$. In Fig.\ref{fig3}b) we report a typical time evolution of $N_{tr} =\sum_{i \neq 0} (N_{+i}+N_{-i})$ in unit of the total number $N_{tot}$ at each time. We observe a clear oscillation with the average amplitude reducing on a timescale of $\approx$ 1 s. To our knowledge the observed dynamics is characterized by the longest coherence time reported in the letterature for trapped condensates separated by a distance of tens of microns \cite{Shin, schumm2005matter, Jo, Hinds, berrada2013integrated}. We think that these results already prove the high intrinsic stability of a BNSL. The observed decay of the oscillation contrast might be due the presence of a spurious harmonic potential of the order of 1 Hz along the \textit{x} axis. Another source of decoherence could be a residual interaction energy due to a non perfect control of the collisional scattering lenght a the level of 0.05 $a_0$, where $a_0$ is the Bohr radius. Bloch oscillations using more homogeneous forces and experiments performed with variable atom numbers will allow to identify what is currently limiting the performance of our Bloch oscillations interferometer. 
\begin{figure}[t]
\centerline{\includegraphics[width=0.85\columnwidth]{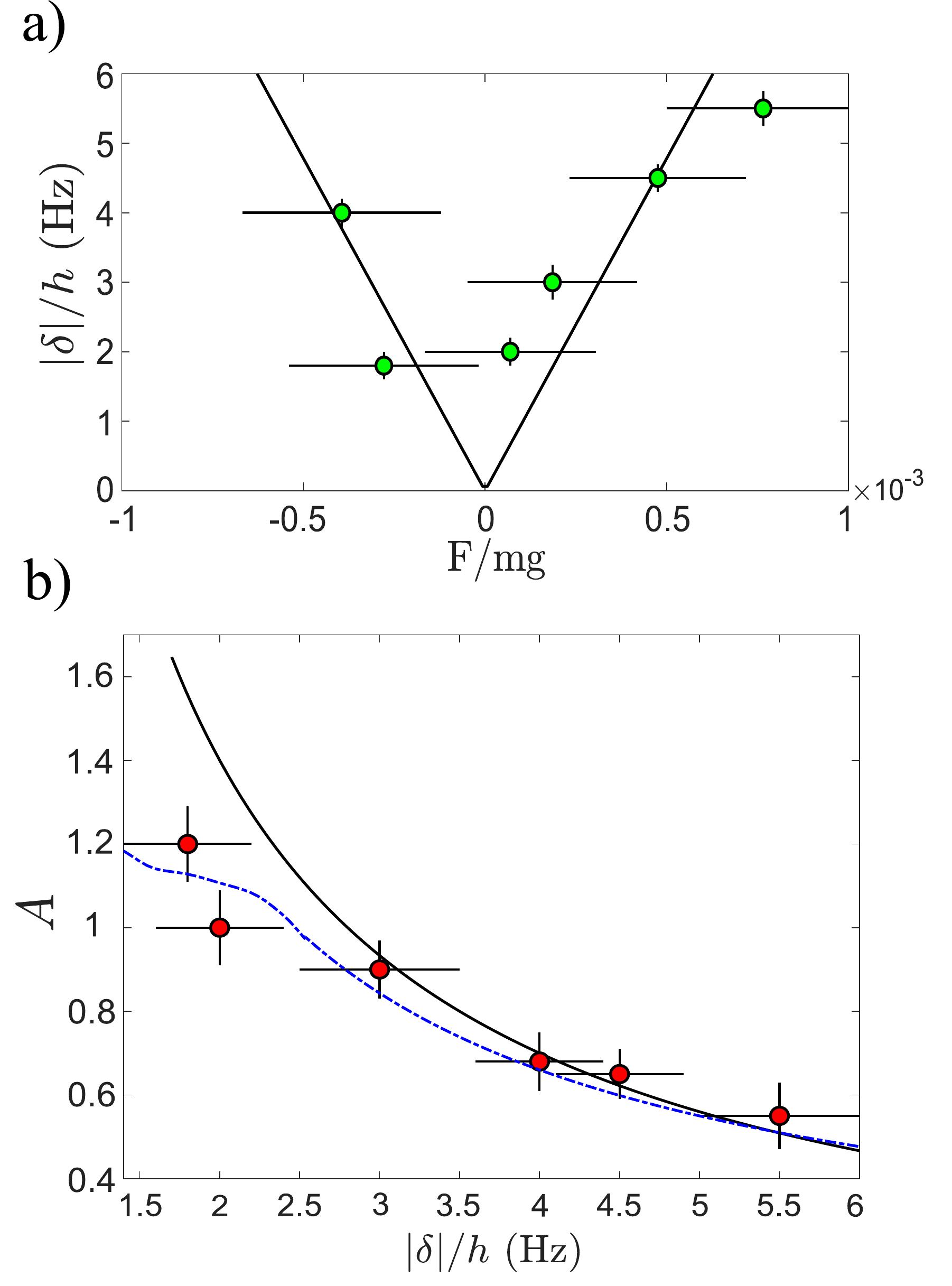}}
\caption{a) Energy difference between lattice sites (dots) derived from the Bloch frequency measurements as a function of the external force in unit of the gravity force $mg$. The horizzontal error bars derive from the force calibration error \cite{SI}. The solid line is the theoretical prediction for a perfect linear potential. b) Amplitude of the oscillation $A$ as a function of the energy difference between lattice sites. The error bars correpond to the uncertainty of the amplitude and the frequency provided by the fit of the oscillations. The solid line correponds to $A=4J_{eff}/\delta$. At lower values of $\delta$ the effect of force inhomogeneities are more significant and both frequency and amplitude deviate from theory. The dashed line is the result of numerical simulations where a longitudinal harmonic potential of 1.5 Hz is included. }
\label{fig4}
\end{figure}\\
We fit the data with a phenomenological model that includes Eq. (\ref{bessel}) and an exponential decay of the amplitude toward a steady state population \cite{SI}. From the fit we can estimate the Bloch period $t_B$, that is related to the external force via the relation $t_B = 1/\delta = h/(F d_-)$, and the amplitude of oscillation $A$, that is linked to the tunneling energy via the equality $A=4J_{eff}/\delta$ (see Eq.\ref{bessel}). We finally investigate the linear dependence of $\delta$ on the external force $F$, calibrated as a function of the dipole trap power perfoming time of flight experiments (see \cite{SI}). \color{black} Results are reported in Fig.\ref{fig4}a) where the line indicates the expected result. Note that only a limited range of forces has been explored because outside it, the small depth $V_{eff} \approx 2$ nK implemented, would cause Landau-Zener interband transitions and atom losses. This could be prevented increasing $V_0$, but the consequent reduction of the tunneling energy $J_{eff}$ would reduce also $A$ and the visibility of the spatial oscillations.  In Fig.\ref{fig4}b) we finally plot the amplitude $A$ as a function of $\delta$ and compare the results with the values we expect from the estimation $J_{eff}=0.7(\pm0.05$) Hz, derived from the experimental calibration of $V_0$. We note that a good agreement is achieved in both plots, although a deviation of the measurements from the expected values is observed for Bloch frequencies $\approx 2$ Hz. For small values of the force, precise measurements are currently limited by the decoherence sources described above.\\
In conclusion, we report the realization of an innovative potential where we exploit the beat-note between two retroreflected laser beams with commensurate wavelengths. Our studies prove that the resulting potential is equivalent to a large spacing single wavelength optical lattices in the limit of small depths, i.e. $V_0 < E_{R^+}$. For larger values of $V_0$, even if the analogy is no longer valid, BNSLs can be used like large spacing standard lattices to create arrays of atomic ensambles with negligible tunneling between the spatial modes. The high stability of a BNSL is demonstrated observing the longest coherent evolution ever reported for a BEC trapped in spatial modes separated by tens of microns. We expect that BNSLs will strongly contribute to the improvement of the sensitivity of trapped atom interferometers \cite{Nzalec} and will represent a valuable tool for the precise manipulation of atoms at large distances in several future quantum technologies \cite{Hosten,trenkwalder2016quantum,Browaeys2020ManybodyPW}.     
\color{black}
\begin{acknowledgments}
We thank all our colleagues of the Quantum Degenerate Group at LENS for support, M. Modugno for inspiring discussions during the early stages of this study and D. Trypogeorgos for careful reading of the manuscript. This work was supported by the project TAIOL of QuantERA ERA-NET Cofund in Quantum Technologies (Grant Agreement No. 731473) implemented within the European Union’s Horizon 2020 Programme.
\end{acknowledgments}

%

\end{document}